# Strong electron – electron correlation and weak localization in CeO$_{0.9}$F$_{0.1}$Fe$_{1-x}$Co$_x$As

S.J. Singh[a], J. Prakash[b], S. Patnaik[a]* and A.K. Ganguli[b]*

[a]School of Physical Sciences, Jawaharlal Nehru University, New Delhi 110067 India
[b]Department of Chemistry, Indian Institute of Technology, New Delhi 110016 India

PACS 47.32.C- Vortex dynamics

PACS 74.25. Fy Transport properties (electric and thermal conductivity, thermoelectric effects, etc.)

PACS 74.62. Dh Effects of crystal defects, doping and substitution

## Abstract

Electron-doping of the semimetal (CeOFeAs) by either fluorine (max $T_c \sim 43$ K) or cobalt (max $T_c \sim 11$ K) leads to superconductivity. Here we show the effect of transition metal (Co) substitution at the iron site on the superconducting properties of CeO$_{0.9}$F$_{0.1}$FeAs ($T_c \sim 38$ K) to understand the interplay of charge carriers in both the rare earth – oxygen and Fe-As layers. Simultaneous doping of equivalent number of charge carriers in both layers leads to a $T_c$ of 9.8 K which is lower than the $T_c$ obtained when either the conducting layer (CeAs) or charge reservoir layer (CeO) is individually doped. This suggests a clear interplay between the two layers to control the superconductivity. The resistivity shows a $T^2$ dependence (T $\gg T_c$) which indicates strong electron-electron correlation. Hall coefficient and thermoelectric power indicate increased carrier concentration with cobalt doping in CeO$_{0.9}$F$_{0.1}$FeAs. The rf penetration depth both for CeO$_{0.9}$F$_{0.1}$Fe$_{0.95}$Co$_{0.05}$As and CeO$_{0.9}$F$_{0.1}$FeAs show an exponential temperature dependence with a gap value of ~ 1.6 and 1.9 meV. A resistance minimum is observed in the normal state near $T_c$ which also shows negative magnetoresistance and provides evidence for the onset of weak localization.

*Authors for correspondence
[b]E-mail: - ashok@chemistry.iitd.ernet.in
[a]E-mail: - spatnaik@mail.jnu.ac.in

# Introduction

Superconductivity in the ferropnictides has provided a novel opportunity for a deeper understanding of the mechanism of high temperature superconductivity in the presence of competing magnetic and superconducting correlations. Consequently, there have been several efforts to draw a parallel between the high $T_c$ cuprates and ferropnictides [1-2]. It has now been established that unlike cuprates, where the ground state is a Mott Insulator with superexchange linked local moments at copper sites, the oxypnictides are (semi)metallic with effectively very weak magnetism as evidenced by a spin density wave correlation. Band structure calculations and photoelectron spectroscopic studies have revealed itinerant characteristics of Fe 3d electrons [2]. Further, amongst the ferropnictides, the Ce(O/F)FeAs (maximum $T_c \sim 43$ K) system is unique because of striking similarity [3] of the electronic phase diagram with cuprates, notably a dome-shaped $T_c$ dependence on carrier doping [4]. The parent oxypnictide, CeOFeAs, undergoes a structural phase transition from tetragonal ($P4/nmm$) to orthorhombic ($Cmma$) structure around 155 K followed by an antiferromagnetic ordering of spins in the Fe sublattice at ~ 140 K. From detailed neutron scattering measurements it is now established that upon F doping in CeO$_{1-x}$F$_x$FeAs, both AFM ordering and the structural phase transition vanishes at x> 0.06 and x~0.1 respectively leading to superconductivity with a maximum $T_c$ (~40 K) [4]. Most strikingly, it has been shown that instead of F at O sites, Co doping at Fe sites in LnOFeAs can also induce superconductivity [5, 6]. This is surprising because LaOCoAs is a metal that becomes a ferromagnet at 66 K [7]. All the above properties indicate that magnetic fluctuations play a critical role in the stabilization of the superconducting state in ferropnictides. To get further insight into the electronic correlation effects of the FeAs layer, we have chosen to dope the iron sites with cobalt in the superconducting composition ($CeO_{0.9}F_{0.1}FeAs$) where the AFM (SDW) state has already been suppressed by F doping. In the undoped CeOFeAs, the nearest neighbor and next nearest neighbor interactions are superexchange controlled and lead to a frustrated stripelike AFM state [5]. Cobalt doping at Fe sites lead to a double exchange interaction between Fe and Co sites. Thus the role of Co doping in optimally doped $CeO_{0.9}F_{0.1}FeAs$ is two fold; to over-dope carriers onto the FeAs (conducting) layers and to introduce disorder in the Fe antiferromagnetic order (dormant due to F doping). While our Hall and thermoelectric power data clearly indicate increase in carrier concentration with Co doping, the normal state shows increased resistivity with clear signature of weak localization effects.

## Experimental

Polycrystalline samples with nominal compositions of $CeO_{0.9}F_{0.1}Fe_{1-x}Co_xAs$ with 'x' = 0.05, 0.10 and 0.15 were synthesized by a two step solid state reaction method [8] using high purity Ce, $CeO_2$, $CeF_3$, $Co_3O_4$, As and FeAs as starting materials. FeAs was obtained by reacting Fe chips and As powder at 800 ºC for 24 hours. The raw materials (all with purities better than 99.9 %) were taken according to stoichiometric ratio and then sealed in evacuated silica ampoules ($10^{-4}$ torr) and heated at 950 °C for 24 hours. The powder was then compacted (5 tonnes) and the disks were wrapped in Ta foil, sealed in evacuated silica ampoules and heated at 1150 ºC for 48 h. All chemicals were handled in a nitrogen-filled glove box. The resulting samples were characterized by powder x-ray diffraction (PXRD) with Cu-$K\alpha$ radiation. The lattice parameters were obtained using a least squares fit to the observed $d$ values.

Resistivity, Hall effect and rf susceptibility measurements were carried out using a cryogenic 8 T cryogen-free magnet in conjunction with a variable temperature insert (VTI). The samples were cooled in helium vapor and the temperature was measured with an accuracy of 0.05 K using a calibrated Cernox sensor wired to a Lakeshore 340 temperature controller. Standard four probe technique was used for transport measurements. The external magnetic field (0-3 T) was applied perpendicular to the probe current direction and the data were recorded during the warming cycle with heating rate of 1 K/min. The inductive part of the magnetic susceptibility was measured using a tunnel diode based rf penetration depth technique [9]. The sample was kept inside an inductor that formed a part of an LC circuit of an ultrastable oscillator (~2.3MHz). A change in the magnetic state of the sample results in a change in the inductance of the coil and is reflected as a shift in oscillator frequency which is measured by an Agilent 53131A counter. The thermo electric power data were obtained in the temperature range of 14 K to 300 K in the bridge geometry across a 2 mm by 3 mm rectangular disk.

**Results and discussion**

Figure 1(i) shows the PXRD patterns of $CeO_{0.9}F_{0.1}Fe_{1-x}Co_xAs$ (x = 0.05, 0.1 and 0.15) compounds. The observed reflections could be satisfactorily indexed on the basis of a tetragonal ZrCuSiAs type structure. Figure 1 (ii) shows the variation of lattice parameters as a function of Co content. With increasing Co content (x), the '**a**' axis goes down marginally while the **c**-axis shrinks significantly indicating successful incorporation of cobalt ions (the ionic size of cobalt(II) is smaller than that of iron(II)).

The zero field resistivity plots between 1.6 K and 300 K for $CeO_{0.9}F_{0.1}Fe_{1-x}Co_xAs$ (Fig. 2) show metallic behavior in the normal state. The sample without Co doping ($CeO_{0.9}F_{0.1}FeAs$) shows the superconducting transition at around 38.5 K [10]. With Co doping, the onset of superconducting transition decreased to 23.4 and 9.8 K for (a) $CeO_{0.9}F_{0.1}Fe_{0.95}Co_{0.05}As$ and (b) $CeO_{0.9}F_{0.1}Fe_{0.9}Co_{0.1}As$ respectively (Fig. 2). The diamagnetic behavior of these superconductors was confirmed from the susceptibility studies which are shown in Fig. 7. On increasing the Co-content (x = 0.15) further the $T_c$ decreases dramatically and only a steep drop in the resistivity around 4 K (Fig 3d) was observed (zero resistance state could not be achieved down to 1.6 K). The residual resistivity value (RRR = $\rho_{300} / \rho_{25}$) is 5.8 and 2.11 for x = 0.05 and 0.1 compositions respectively (RRR = 4.34 for $CeO_{0.9}F_{0.1}FeAs$ [10]) indicate increase in disorder with Co doping. It is to be noted that while in the compositions without fluorine $CeOFe_{1-x}Co_xAs$, both the transition temperature and $\rho(T_c)$ always decreased with increasing x > 0.01 [6], on the contrary we observe that $\rho(T_c)$ first increases with Co doping but reduces drastically for x ~ 0.15 and shows dome like behavior(optimal dopant concentration exists). This is reminiscent of distinctly different transport behavior in the metallic regions of $LaO_{1-x}F_xFe_{1-y}Co_yAs$(x= 0.1) [11]. Various studies have shown that optimal $T_c$ in all oxypnictide superconductors depends on the presence of ideal (undistorted) $FeAs_4$ tetrahedra (bond angle of 109° 28'). In $CeO_{0.9}F_{0.1}FeAs$, this angle is found to be ~ $112^0$ [4], thus with Co doping the decrease in $T_c$ may be related to increase in the Fe-As-Fe angle though the interlayer distance between CeO/F and FeAs layers decreases by doping Co in place of Fe. Our results are in line with what is seen in LaO(Fe/Co)As, where the bond angle increases with Co doping, leading to a maximum $T_c$ at Fe-As-Fe ~ $113.5^0$. This reinforces the understanding that Fe-As-Fe bond angles play a dominant role in determining the optimal $T_c$, more than the decreasing distances between the reservoirs and conducting layers.

We next discuss the striking increase in $\rho(T_c)$ with cobalt doping for x < 0.15. Figure 3 shows the evolution of the temperature dependence of normal state resistivity as a function of Co doping. To establish whether the carrier concentration increased or decreased with Co doping, we have investigated Hall coefficient (Fig. 4) and thermoelectric power data (Fig. 5) respectively. In Fig. 4 (a) we show the Hall resistivity ($\rho_{xy}$) for all three Co doped compounds ( x= 0.05, 0.1 and 0.15) and compare it with $CeO_{0.9}F_{0.1}FeAs$ (x=0). For a normal metal the variation of Hall coefficient ($R_H$) with temperature remains nearly invariant with temperature. However, the Hall coefficient varies with temperature for a multiband material such as $MgB_2$ [12] or materials with non-Fermi liquid behavior such as the cuprates [13]. As shown in Fig 4(a), we find that the transverse resistivity ($\rho_{xy}$) remains negative upto room temperature for all the superconductors being discussed here. This indicates that the transport is dominated by the electrons and we find that the slope of $\rho_{xy}$ vs. H decreases with increasing cobalt concentration. This is a clear proof that carrier concentration increases with substitution of cobalt in place of Fe in this optimally doped superconductor $CeO_{0.9}F_{0.1}FeAs$. Further, in Fig. 4(b) we find that Hall coefficient ($R_H$) for x = 0 composition shows strong temperature dependence in comparison to Co doped x = 0.05 compound which suggests a relatively stronger multiband effect in $CeO_{0.9}F_{0.1}FeAs$. However, it should be pointed out that the Hall coefficient ($R_H$) cannot be simply expressed as 1/ne for a multiband superconductor and these conclusions are approximate.

The temperature dependence of the Seebeck coefficient (S) for CeO$_{0.9}$F$_{0.1}$Fe$_{1-x}$Co$_x$As is shown in Fig. 5. In the measured temperature range, S has a negative value which is similar to that found in the parent Ce(O/F)FeAs [10] and again confirms dominant electronic conduction [8]. For CeO$_{0.9}$F$_{0.1}$FeAs (without Co doping), the Seebeck coefficient varies from -22 µV / K at 300 K to a value of ~ - 39 µV / K at ~ 86 K and then decreases sharply (in magnitude) as the temperature is lowered further. The above data indicates that the absolute value of thermopower decreases with increasing Co-doping. In a minimal two band model, the conductivity and thermopower depend on hole and electron contributions, and therefore while it is certain that with cobalt doping the conduction mechanism is dominated by electrons, the relative decrease in absolute value of S is due to the contribution from holes in the over-doped regime. Further, from the behavior of the temperature dependence of resistivity, we find that ρ above $T_c$ exhibits a quadratic temperature dependence (ρ = A + BT$^2$) between 30 and 200 K for x = 0.05 compound (inset of Fig. 5). The T$^2$ behavior of ρ below 200 K indicates relatively strong electronic correlations and is consistent with the formation of a Fermi-liquid state. The slope B obtained from the fit is 0.94 × 10$^{-4}$ mΩ cm K$^{-2}$ for H = 0 and A = 1.8 mΩ cm. Similarly for x =0.1 compound, B is 0.45 × 10$^{-4}$ mΩ cm K$^{-2}$ and A = 2.86 mΩ cm. This value of the slope (B) is larger than that observed in some semi-heavy fermion compounds [14].

After establishing the addition of carriers with cobalt doping in the over-doped regime, we next focus on the curious resistivity upturn near $T_c$ for Co doped samples the onset of which shifts to lower temperature with higher cobalt concentration. This is shown in Fig 3a –d with an arrow mark indicating resistivity minima for samples with different Co concentration. An explanation for resistance minimum may be sought in the Kondo scattering from dilute magnetic scattering. If this is the case, we would expect logarithmic temperature dependence below $T_{min}$ which does not occur in our case. A disorder induced Anderson localization could also explain the upturn but that would lead to a stronger upturn with increasing Co concentration which contradicts our experimental observation. In the inset of figure Fig. 3c, we plot the magnetic field dependence of resistivity of x = 0.1 composition and clearly observe negative magnetoresistance. Such behavior is indicative of weak localization where the external field, by introducing phase difference between the oppositely traversed loop, reduces the tendency of localization. However this needs to be confirmed by further examining the field dependence of normal state magnetoresistance [15].

To obtain the upper critical field ($H_{c2}$), we have compared the temperature dependence of the resistivity under varying magnetic fields for $CeO_{0.9}F_{0.1}FeAs$ and $CeO_{0.9}F_{0.1}Fe_{0.95}Co_{0.05}As$. From Fig. 6 it is clear that the $T_c$ (onset) shifts weakly with magnetic field, but the zero resistivity temperature shifts much more rapidly to lower temperatures. Using a criterion of 90 % and 10 % of normal state resistivity ($\rho_n$), the upper critical field ($H_{c2}$) and the irreversibility field H*(T) were calculated. The H-T phase diagram for each sample is shown in inset of Fig 6. The zero field upper critical field $H_{c2}(0)$ was calculated using the Werthamer-Helfand- Hohenberg (WHH) formula [16]. Using the value of transition temperature ($T_c$) of 23.4 K, we find $H_{c2}$ = 25.3 T. This value is smaller than the reported $H_{c2}$ value (~94 T) of $CeO_{0.9}F_{0.1}FeAs$ [10]. Using the value of $H_{c2}(0)$ we can also calculate the mean field Ginzberg-Landau coherence length

$(\xi_{ab} = \left(\phi_0 / 2\pi H_{C2}\right)^{1/2}$. Using $\Phi_0$ = 2.07 × $10^{-7}$ Gcm$^2$ and the H$_{c2}$ values, we obtain a coherence length of ~ 36 Å for CeO$_{0.9}$F$_{0.1}$Fe$_{0.95}$Co$_{0.05}$As composition. This value is higher than that reported for sample CeO$_{0.9}$F$_{0.1}$FeAs [10] and indicates that impurity scattering in the superconducting state does not increase with Co doping.

In Fig. 7, we compare the temperature dependence of the penetration depth (Δλ) measured using tunnel diode oscillator technique for sample x = 0 and 0.05 compositions. This dependence is directly related to the anisotropy of the superconducting energy gaps. Both in the fluorine-doped or oxygen-deficient FeAs based superconductors, the majority of earlier reports indicate a fully gapped Fermi surface (FS). Measurements of the London penetration depth, λ(T), using a tunnel diode technique on NdFeAsO$_{0.9}$F$_{0.1}$ [18], La$_{0.8}$Th$_{0.2}$OFeAs [19] and SmFeAsO$_{1-x}$F$_y$ [20] have found an exponential temperature dependence of λ(T) at low temperatures. Further, it has been established that the rf penetration depth technique works better for sintered polycrystalline samples as compared to point contact or tunneling technique [17] because the length scale is an order of magnitude higher. In Fig. 7 we plot the penetration depth for both samples. We have considered penetration depth data up to 0.3 T$_c$ to ensure validity of the low temperature BCS expression for an isotropic s-wave state

$$\Delta\lambda = \lambda(0) \times \sqrt{\frac{\pi\Delta(0)}{2T}} \exp[-\Delta(0)/T] \quad (1)$$

where, Δλ(T) is the difference between penetration depth at the temperature T and at lowest measurement temperature of 1.8 K. λ(0) and Δ(0) are the zero temperature values of penetration depth and energy gap respectively. In the tunnel diode oscillator technique, the change in penetration depth is related to the measured shift in frequency,

$\Delta\lambda = -G\,\Delta f$, where G is a geometrical constant that is calibrated to be ~ 10 for our set up. This was estimated by deriving the skin depth ($\delta = 1/\mu_0\pi\sigma f$, $\sigma$ being the conductivity) from frequency shift for oxygen free high purity copper in a shape roughly similar to the superconducting specimen. We clarify that because of the uncertainty in geometry, the absolute value of penetration depth is only an approximation but the temperature dependence of change in penetration depth is accurate. We fit our data for an isotropic s-wave gap in the range $0 < T/T_c < 0.3$ as shown in the inset of Fig. 7. The solid line in the figure shows a curve fit to an isotropic single gap that yields better fitting (compared to power law dependence) with the gap value of $\Delta_0/k_B = 22.3$ ( gap ~ 1.9 meV) and 18 (gap ~ 1.6 meV ) for $x = 0$ and $x = 0.05$ compositions respectively. Thus the gap value decreases with Co doping. We note that while fitting the data according to Eq. 1 will yield the true estimate of the low lying gap, however the possibility of multiple gaps and anisotropy requires the analysis of the superfluid density ($\lambda^2(0)/\lambda^2(T)$).

## Conclusions

To summarize, we have successfully synthesized cobalt-doped CeO$_{0.9}$F$_{0.1}$Fe$_{1-x}$Co$_x$As (x = 0.05, 0.1, 0.15 and 0.2) superconductors to study the role of charge carriers in the over doped regime. The addition of Co suppresses the transition temperature and upper critical field. While the Hall and thermopower data establish increasing electron dominated conduction mechanism with Co doping, a curious upturn near $T_c$ is observed which suggests the presence of weak localization which is supported by the observation of negative magnetoresistance. Our penetration depth analysis indicates that both CeO$_{0.9}$F$_{0.1}$FeAs and CeO$_{0.9}$F$_{0.1}$Fe$_{0.95}$Co$_{0.05}$As follow S wave pairing symmetry with gap values of 1.9 and 1.6 meV respectively. Away from $T_c$, evidence of strong electronic correlation is established in Co-doped samples from the $T^2$ dependence of resistivity in the normal state. Surprisingly, we see increase in coherence length with increased disorder in the FeAs layer.

**Acknowledgements**

AKG and SP thank DST, Govt. of India for financial support. JP and SJS thank CSIR and UGC, Govt. of India, respectively for fellowships. We thank Prof. Deepak Kumar (JNU) for very useful discussions.

# References:

[1] Mazin I I and Johannes M D 2009 *Nature Physics* **5** 141.

[2] Bondino F, Magnano E, Malvestuto M, Parmigiani F, Mcguire M A, Sefat A S, Sales B C, Jin R, Mandrus D, Plummer E W, Singh D J and Mannella N 2008 *Phys. Rev. Lett*. **101** 267001.

[3] Rao C N R and Ganguli A K 1995 *Chem. Soc. Rev.* **24** 1.

[4] Zhao J, Huang Q, Cruz C D L, Li S, Lynn J W, Chen Y, Green M A, Chen G F, Li G, Li Z, Luo J L, Wang N L and Dai P 2008 *Nature Material* **7** 953.

[5] Wang C, Li Y K, Zhu Z W, Jiang S, Lin X, Luo Y K, Chi S, Li L J, Ren Z, He M, Chen H, Wang Y T, Tao Q, Cao G H and Xu Z A 2009 *Phys. Rev. B* **79** 054521.

[6] Prakash J, Singh S J, Patnaik S and Ganguli A K 2009 *Solid State Commun.* **149** 181.

[7] Yanagi H, Kawamura R, Kamiya T, Kamihara Y, Hirano M, Akamura T, Osawa H and Hosono H 2008 *Phys. Rev. B* **77** 224431.

[8] Sefat A S, Mcquire M A, Sales B C, Jin R, Howe J Y and Mandrus D 2008 *Phys. Rev. B* **77** 174503.

[9] Patnaik S, Singh K J and Budhani R C 1999 *Rev. Sci. Instrum.* **70** 1494.

[10] Prakash J, Singh S J, Patnaik S and Ganguli A K 2009 *Physica C* **469** 82.

[11] Lee S C, Kawabata A, Moyoshi T, Kobayashi Y and *Sato* M, arxiv:0812.3949 (2008).

[12] Yang H, Liu Y, Zhuang C, Shi J, Yao Y, Massidda S, Monni M, Jia Y, Xi X, Li Q, Liu Z-K, Feng Q and Wen H-H 2008 *Phys. Rev. Lett.* **101** 067001.

[13] Ong N P, in Physical Properties of High Temperature Superconductors, Ed. D. M.

Ginsberg (World Scientific, Singapur,1990), p.459.

[14] Kadowaki K and Woods S B 1986 *Solid State Commun.* **58** 507.

[15] Altshuler B L and Aronov A G, Electron-Electron Interactions in Disordered systems ed. EFROS A. L. and POLLAK M. (Amsterdam: Elsevier) (1985) p1.

[16] Werthamer N R, Helfand E and Hohenberg P C 1966 *Phys. Rev.* **147** 295.

[17] Shan L, Wang Y, Zhu X, Mu G, Fang L and Wen H –H 2008 *Europhys. Lett.* **83** 57004.

[18] Prozorov R, Tanatar M A, Gordon R T, Martin C, Kim H, Kogan V G, Ni N, Tillman M E, Bud'ko S L and Canfield P C arXiv:0901.3698 (2008).

[19] Prakash J, Singh S J, Patnaik S and Ganguli A K 2009 *J. Phys.: Condens. Matter* **21** 175705.

[20] Malone L, Fletcher J D, Serafin A, Carrington A, Zhigadlo N D, Bukowsk I Z, Katrych S and Karpinski J 2009 *Phys. Rev. B* **79** 140501.

.

**Figure Caption :**

**Figure: 1. (i)** Powder X-ray diffraction patterns (PXRD) of $CeO_{0.9}F_{0.1}Fe_{1-x}Co_xAs$ (x = 0.05, 0.1 and 0.15) **(ii)** The variation of the lattice parameters (**a** and **c**) on the Co content (x) for $CeO_{0.9}F_{0.1}Fe_{1-x}Co_xAs$.

**Figure: 2.** The temperature dependence of resistivity ($\rho$) as a function of temperature for $CeO_{0.9}F_{0.1}Fe_{1-x}Co_xAs$ (x = 0, 0.05, 0.10 and 0.15). Inset shows the dependence of transition temperature ($T_c$) with Co content (x).

**Figure 3.** The resistivity variation with zero applied field for (a) $CeO_{0.9}F_{0.1}FeAs$, (b) $CeO_{0.9}F_{0.1}Fe_{0.95}Co_{0.05}As$, (c) $CeO_{0.9}F_{0.1}Fe_{0.9}Co_{0.1}As$ and (d) $CeO_{0.9}F_{0.1}Fe_{0.85}Co_{0.15}As$. Arrows show the minimum point ($T_{min}$). The inset figure shows the resistivity variation with temperature in the presence of magnetic field for $CeO_{0.9}F_{0.1}Fe_{0.9}Co_{0.1}As$ (sample c).

**Figure 4. (a)** Hall resistivity as a function of field and **(b)** Hall coefficient as a function of temperature for $CeO_{0.9}F_{0.1}FeAs$ (■) and $CeO_{0.9}F_{0.1}Fe_{0.95}Co_{0.05}As$ (●). The straight line in figure 4a is a guide to eye and shows linearity.

**Figure 5.** Temperature dependence of thermopower (S) for $CeO_{0.9}F_{0.1}Fe_{0.95}Co_{0.05}As$, $CeO_{0.9}F_{0.1}Fe_{0.9}Co_{0.1}As$ and $CeO_{0.9}F_{0.1}Fe_{0.85}Co_{0.15}As$. Inset shows $T^2$ dependence of resistivity for $CeO_{0.9}F_{0.1}Fe_{0.95}Co_{0.05}As$ in the range, 30 K $\leq$ T $\geq$ 200 K.

**Figure 6.** Temperature dependence of the electrical resistivity of $CeO_{0.9}F_{0.1}Fe_{0.95}Co_{0.05}As$ with varying magnetic fields. Inset shows the temperature dependence of upper critical field (■) and irreversibility field (●) as a function of temperature.

**Figure 7.** Temperature dependence of penetration depth for (a) $CeO_{0.9}F_{0.1}FeAs$ and (b) $CeO_{0.9}F_{0.1}Fe_{0.95}Co_{0.05}As$ upto the transition temperature attesting to the onset of bulk diamagnetism at $T_c$. The inset shows variation of penetration depth, $\Delta\lambda$ (T) for both the samples. The red and blue lines show the exponential fitting.

**Figure 1:**

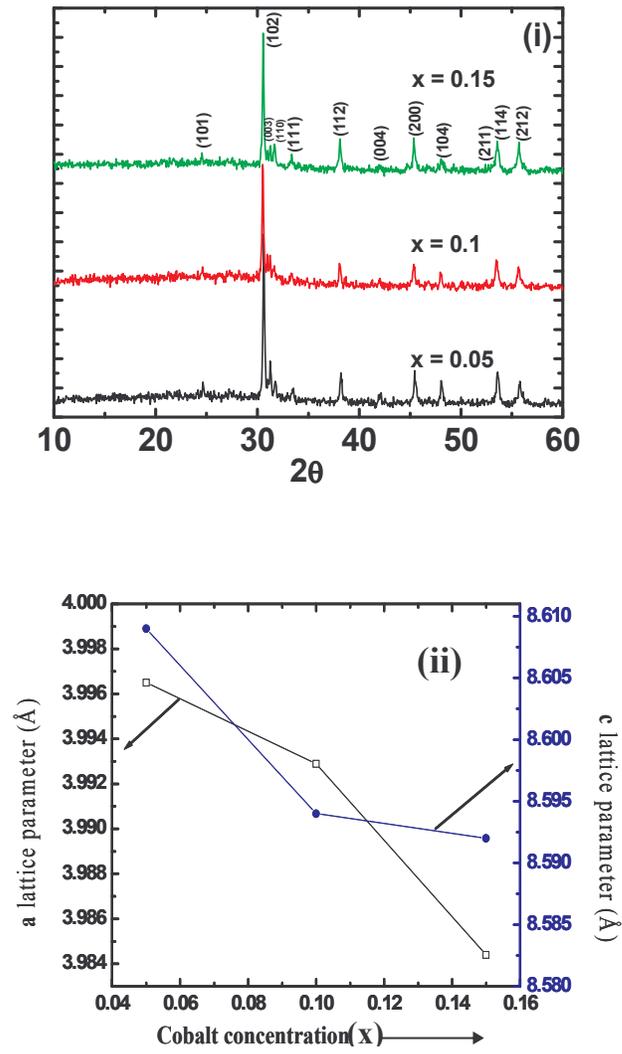

**Figure 2.**

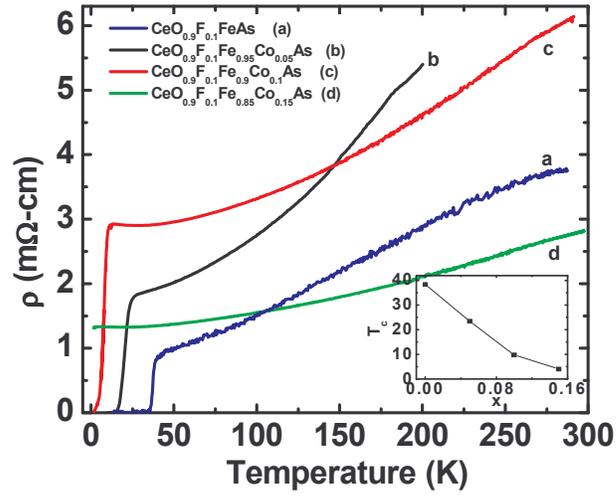

**Figure 3.**

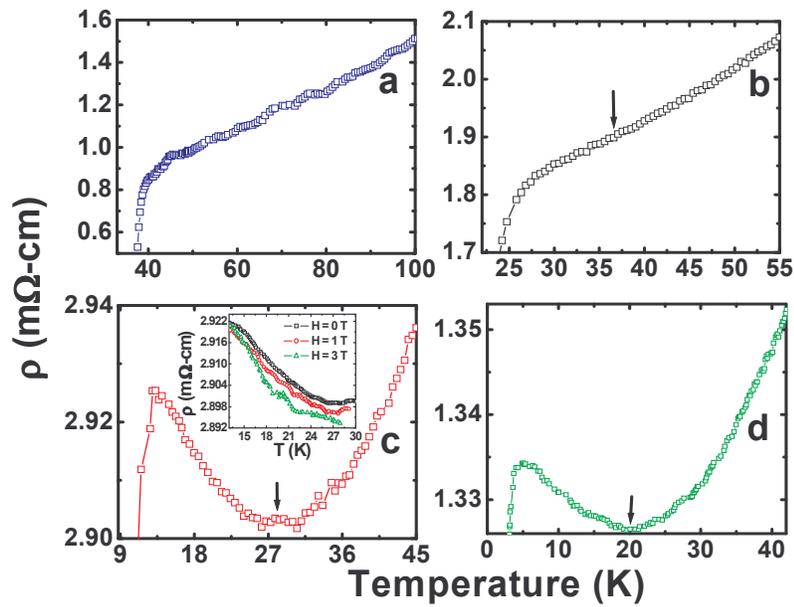

**Figure 4 :**

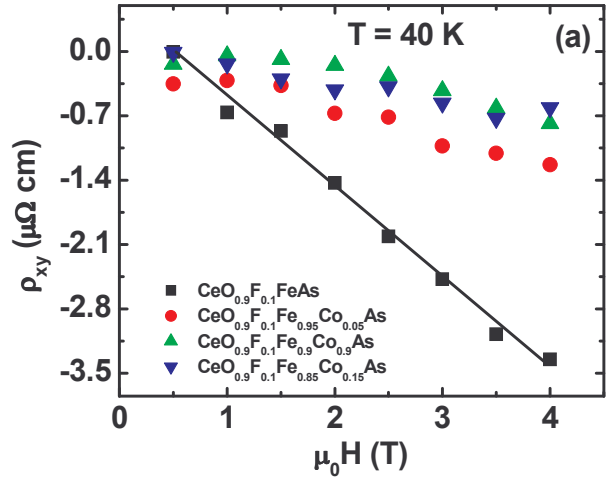

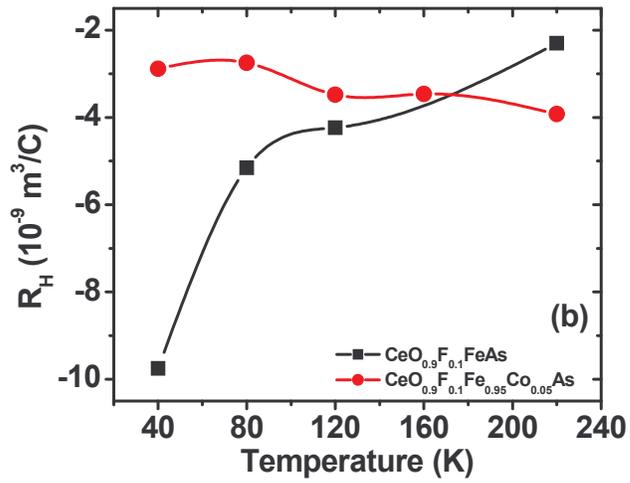

**Figure 5.**

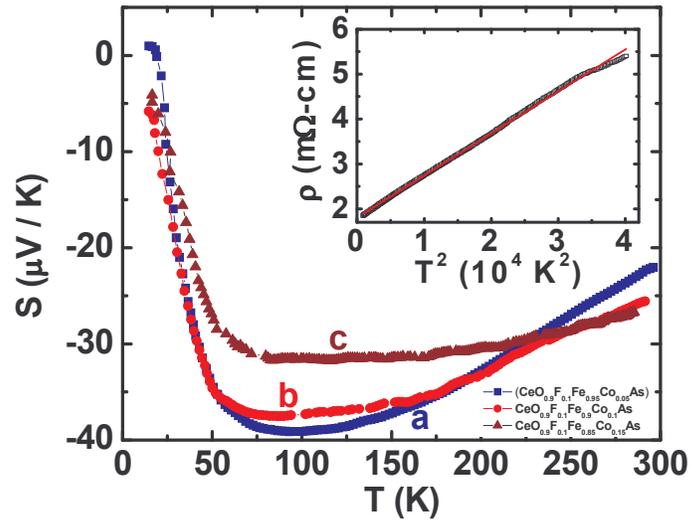

**Figure 6 :**

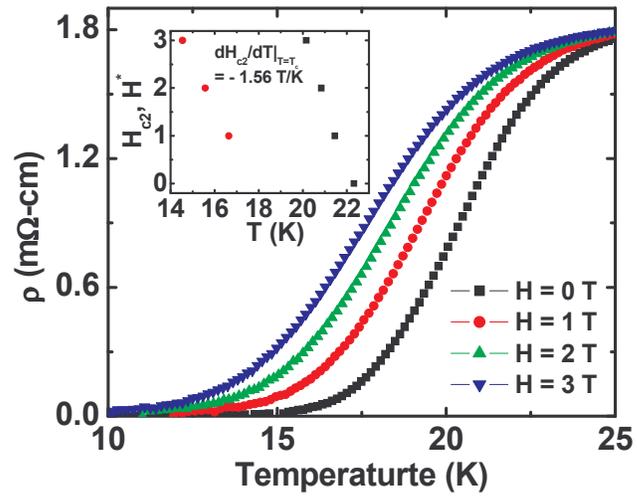

**Figure 7 :**

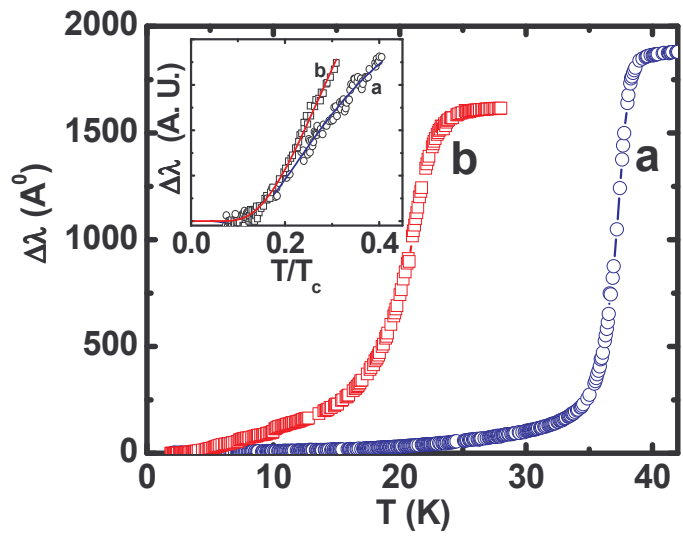